\documentclass[runningheads]{llncs}
\usepackage{graphicx}
\usepackage{tabularx}
\usepackage{makecell}
\begin{document}
\title{Tracking environmental policy changes in the Brazilian Federal Official Gazette}
%
\titlerunning{Tracking changes in the Brazilian Federal Official Gazette}
%
\author{Flávio Nakasato Cação\thanks{Author to whom correspondence should be addressed.}\inst{1} \orcidID{0000-0003-4771-6009} \and
        Anna Helena Reali Costa\inst{1} \orcidID{0000-0001-7309-4528} \and
        Natalie Unterstell\inst{2} \orcidID{0000-0002-4340-4916} \and
        Liuca Yonaha\inst{2} \orcidID{0000-0002-1122-4435} \and
        Taciana Stec\inst{2} \orcidID{0000-0001-7911-143X} \and
        Fábio Ishisaki\inst{2} \orcidID{0000-0001-9839-7699}}
        
\authorrunning{F. Cação et al.}

\institute{Escola Politécnica, Universidade de São Paulo, Sao Paulo, Brazil \\
\and Política por Inteiro, Sao Paulo, Brazil \\
\email{\{flavio.cacao, anna.reali\}@usp.br} \\
\email{\{natalie, liuca, taciana, fabio\}@politicaporinteiro.org}}

\maketitle              
\begin{abstract}
Even though most of its energy generation comes from renewable sources, Brazil is one of the largest emitters of greenhouse gases in the world, due to intense farming and deforestation of biomes such as the Amazon Rainforest, whose preservation is essential for compliance with the Paris Agreement. Still, regardless of lobbies or prevailing political orientation, all government legal actions are published daily in the Brazilian Federal Official Gazette (BFOG, or ``Diário Oficial da União" in Portuguese). However, with hundreds of decrees issued every day by the authorities, it is absolutely burdensome to manually analyze all these processes and find out which ones can pose serious environmental hazards. In this paper, we present a strategy to compose automated techniques and domain expert knowledge to process all the data from the BFOG. We also provide the Government Actions Tracker, a highly curated dataset, in Portuguese, annotated by domain experts, on federal government acts about the Brazilian environmental policies. Finally, we build and compared four different NLP models on the classfication task in this dataset. Our best model achieved a F1-score of $0.714 \pm 0.031$. In the future, this system should serve to scale up the high-quality tracking of all oficial documents with a minimum of human supervision and contribute to increasing society's awareness of government actions.

\end{abstract}
\section{Introduction}
Brazil has one of the largest reserves of biodiversity in the world, such as the Amazon Rainforest, Cerrado and Atlantic forest. The preservation of these biomes is essential for the country to be able to fulfill the objectives of the Paris Agreement \cite{Rochedo2018}, since 78\% of greenhouse gas emissions in Brazil come from land use and cover change \cite{West2019}. In 2020, while global emissions fell as a result of the coronavirus pandemic, in Brazil they grew substantially driven by deforestation and farming \cite{Reuter2020}; the Amazon Rainforest deforestation rate was the greatest of the decade \cite{SilvaJunior2021}. At the same time, the country is an agribusiness powerhouse, with 26.6\% of the GDP related to it \cite{Esalq2020}. Despite this complex and dynamic environment, governments at all hierarchical levels of the nation are required to record all their legal actions in the Official Gazette. Thus, one can infer potentially harmful directions for the environmental policy from the systematic scrutiny of these documents. This makes tracking government acts a powerful tool to alert journalists and empower civil society with qualified and clear information \cite{Rolnick2019}. However, this is an arduous task for manual work alone \cite{Grimmer2013}: hundreds of highly technical documents are issued every day by the Legislative and the Executive branches, at the federal, state and municipal levels. According to the Brazilian Institute of Geography and Statistics (IBGE), Brazil has 5570 municipalities and 26 states in addition to its Federal District\footnote{IBGE: \url{https://cidades.ibge.gov.br/brasil/panorama}.}, and typically each of these federative entities has its own Official Gazette.

Therefore, this scenario represents a yet underexplored opportunity for the most recent pre-trained language models, particularly for the Brazilian Portuguese language. Pretrained language models, such as BERT \cite{Devlin2018} and T5 \cite{JMLR:v21:20-074}, have started to become popular in recent years and have set new quality standards in virtually all natural language processing (NLP) tasks, such as classification, translating and question answering. They have millions or billions of parameters and are built upon the Transformer architecture \cite{Vaswani2017}, which leverages self-attention mechanisms and eliminates the need for recurrent neural networks. This allows models to be trained in parallel in a self-supervised way over huge databases, such as the entire Wikipedia. Finally, the parametric knowledge accumulated in its parameters can thus be transferred efficiently to general language tasks after minor fine-tunings on smaller domain-specific datasets. The results obtained in this way usually outperform models trained solely on the aforementioned smaller dataset \cite{JMLR:v21:20-074}. 

Work presented in \cite{stede2021climate} draws attention to the existing opportunities for collaboration between the NLP community and social scientists in studies related to climate change, such as tracing political discourses, topic modeling, and extracting insights when there is not ``Big Data" -- which is often a prerequisite in machine learning. \textit{ClimateQA} is one of the most recent initiatives to tackle public and open data with advanced NLP models \cite{Luccioni2020}. It consists of a RoBERTa model \cite{liu1907roberta} that can be adjusted in any company's reporting database to answer questions about sustainability by consulting the company's unstructured files. Other works, such as \cite{hatonen2021from}, address the lack of accountability of politicians by providing a topic modeling system to aggregate policy makers' speeches from multiple data sources, such as Twitter and Facebook. Such systems have the potential to help the public more clearly discern opinions that are currently in vogue in public debate.

In the Portuguese language, the scarcity of resources, datasets or models, is even more striking. Among the initiatives, DEBACER is an algorithm developed to automatically segment blocks of speeches by politicians registered in the minutes of the Portuguese Parliament \cite{ferraz2021debacer}. On the environmental side, Pirá is the first Portuguese-English bilingual question answering (QA) dataset on the Brazilian coast and oceans in general; it was crowdsourced and contains 2,261 question answer pairs on these subjects \cite{Paschoal2021}. In the same direction, DEEPAGÉ is a QA system dedicated to answer questions about the Brazilian environment in Portuguese. It runs over news and Wikipedia articles on the subject and was fine-tuned on QA pairs filtered and translated from a massive open-domain QA dataset, due to the lack of Portuguese QA datasets on the topic \cite{caccao2021deepage}.

To cover the gap of approaches like those mentioned above in the case of the Portuguese language and address important federal government acts, in this work, we collected thousands of historical documents from the Brazilian Federal Official Gazette (BFOG) -- or, as it is known in Brazil, ``Diário Oficial da União" (DOU) -- and compared multiple NLP models to classify changes in environmental policies. In order to do so, we first built a rule-based robot to scrap all the official documents from the BFOG, filter and pre-classify based on keywords defined by domain experts. Thus, the same domain experts reviewed and enriched a share of this initial data specifically related to environmental issues. Finally, this curated dataset was splitted and used to train and compare multiple NLP models on the classification task of federal government acts. We tested 4 models: from a traditional Naive Bayes and BiLSTMs to two state-of-the-art techniques, based on BERT \cite{Devlin2018}, a bidirectional Transformer encoder architecture. To summarize, our main contributions are:

\begin{itemize}
    \item \textbf{A new approach to collecting and classifying Brazilian Federal Official Gazette data}, that takes advantage of automatic pre-classification techniques and knowledge from domain experts;
    \item The \textit{\textbf{Government Actions Tracker}} (GAT), an ever-increasing highly curated dataset, in Portuguese, of federal government acts related to the main Brazilian environmental policies -- the dataset is made available at \url{https://www.politicaporinteiro.org/monitor-de-atos-publicos/}\footnote{The version of the GAT dataset used specifically in this work can be downloaded from: \url{https://github.com/nakasato/gat}.};
    \item \textbf{Comparison among multiple NLP models}, from LSTMs to BERT, designed to classify the acts aforementioned;
    \item A \textbf{BERT model fine-tuned} on a Masked Language Model (MLM) task over a corpus of 500k raw documents (not included in GAT) from the BFOG. It is made available at \url{https://huggingface.co/flavio-nakasato/berdou\_500k}.
\end{itemize}

It is noteworthy that the system formed only by the rule-based robot followed by a layer of human supervision today feeds one of the largest newspapers in the country with daily monitoring of acts by the Brazilian government that may have negative consequences for the preservation of the country's native forests and wildlife\footnote{The Environmental Policy Monitor can be accessed here: \url{https://arte.folha.uol.com.br/ambiente/monitor-politica-ambiental}.}. Thanks to this, it was possible to identify massive repeals of protection laws moved by the Federal Government in 2020, with the potential increase in deforestation\footnote{A newspaper article reporting this can be found here: \url{https://www1.folha.uol.com.br/ambiente/2020/07/governo-acelerou-canetadas-sobre-meio-ambiente-durante-a-pandemia.shtml}.}. However, since this current process requires the evaluation of human experts, a more effective classification system, such as these presented in this work, could eliminate the need for human supervision in the vast majority of cases, allowing their efforts to be redirected to new challenges, and dramatically scaling the model's tracking capability.

In the next sections, we will cover the construction strategy of the datasets, highlighting the domain experts' role in this effort and how the rule-based robot performed the initial classifications. Thus, we present GAT, the dataset used for training our models, as well as their main settings. Finally, we discuss our results, presenting future work perspectives.

\begin{figure*}[ht]
\vskip 0.2in
\begin{center}
\centerline{\includegraphics[width=1.0\textwidth]{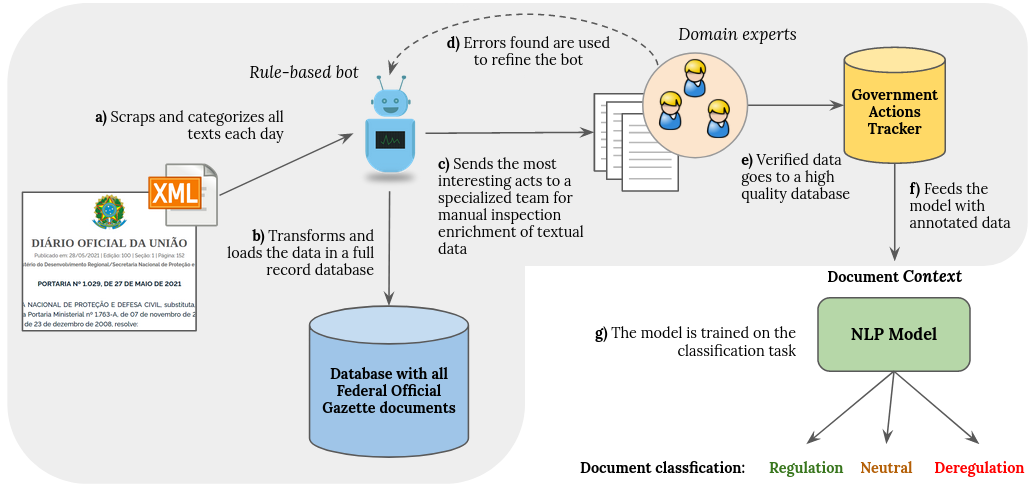}}
\caption{The operational flow of the data pipeline. A rule-based robot scrapes and pre-classifies all official documents released every day (step \textit{a}), and loads them into a general database (step \textit{b}). The most relevant acts related to the environment are sent to a team of domain experts (step \textit{c}), who manually reviews the robot's classifications and enriches the database with new information. Errors found by annotators are regularly used to improve the rule-based robot (step \textit{d}). This filtered and verified subset of data is loaded into the \textit{Government Actions Tracker} database (step \textit{e}), which is used to train the NLP models. These models receive a \textit{Context} variable for each document and are trained/fine-tuned to classify it as a Regulation/Deregulation/Neutral action (step \textit{g}). The system over the gray area represents the Política Por Inteiro's system currently deployed in production.}
\label{data_pipeline}
\end{center}
\vskip -0.2in
\end{figure*}

\section{Methods}

\subsection{Data Preparation}
Every morning, a robot scrapes all documents published in the Federal Official Gazette\footnote{The official documents of the federal government, originally in PDF, are also published in a machine readable format -- in this case, XML. These are the files processed by the robot.} and pre-classifies them under ``Themes'', based on rules defined by domain experts and refined over the years. These rules are based mainly on keywords and more complex expressions to include or exclude a document from a given theme. So far, there are 22 possible themes like \textit{Climate Change}, \textit{Amazon Region} and \textit{Environmental Disasters}, and this pre-tagging helps the robot and domain experts to make an initial filter of the most pertinent documents. All official document data is transformed and loaded into a database.

The most relevant documents filtered by the robot are also sent to a separate file, where, every day, two specialists jointly review them and annotated an \textbf{Action}, a \textbf{Circumstance} and a \textbf{Classification} fields for each record, besides some more useful metadata. An \textbf{Action} refers to the legal action defined by the document, while a \textbf{Circumstance} usually carries more details about the action taken. Both are concatenated into a new variable \textbf{Context} we created to feed the models. Also, for the most part, the two strings are only \textit{extracted from the original document with minimal adjustments}. We expect this might make it more natural to adapt and scale our models to other, new and larger, official documents mentioned in the previous section, which lack human annotators and reviewers. The process is illustrated in Figure \ref{data_pipeline}. Regarding the \textbf{Classification} field, domain experts defined 12 classes, described below:

\begin{itemize}
    \item \textbf{Regulation:} Action that seeks to institute a rule or norm by the public administration, giving guidelines and producing guidance to economic agents;
    \item \textbf{Deregulation:} Action that seeks to revoke and/or reverse a previously established regulation, change its understanding or orientation;
    \item \textbf{Institutional reform:} Change in structure, skills and institutional arrangement related to public policy;
    \item \textbf{Response:} Action that aims to respond to a significant external event, such as a natural disaster or a major accident;
    \item \textbf{Flexibilization:} Alteration, temporary or not, of deadlines or conditions for compliance with environmental rules, norms and legislation;
    \item \textbf{Neutral:} Action with no significant impact when considered in isolation, but cataloging assessed as necessary because it addresses topics on relevant agendas or with indications of becoming relevant in the medium and long terms;
    \item \textbf{Retreat:} Action that seeks to revoke, replace or modify previously established regulations, due to political or popular pressure;
    \item \textbf{Law consolidation:} Result of regulatory review, with no impact on content;
    \item \textbf{Revocation:} Batch revisions or acts associated with the full revision process;
    \item \textbf{Privatization:} Action that seeks the alienation of business rights under the competence of the Union; the transfer, to the private sector, of the execution of public services operated by the Union; or the transfer or grant of rights over movable and immovable property of the Union;
    \item \textbf{Legislation:} Action that seeks to agree a new law before society, giving guidelines and providing guidance to economic agents;
    \item \textbf{Planning:} Action that does not institute regulatory processes per se, but discloses documents and guiding strategies, such as management plans, creation of committees and working groups, approval of programs and policies that have not yet been defined, among others.
\end{itemize}

Misclassifications found by the annotators are also used regularly to refine the rule-based robot, in a process of continuous feedback adjustment via active learning. In the curation process of the GAT dataset, specialists even regularly double-checked the original BFOG documents themselves, which substantially improved the rule-based robot recall over time, minimizing the chances of loss of relevant material.

\begin{table}
\centering
    \caption{\textit{Theme}, \textit{Action} and \textit{Circumstance} and its and respective \textit{Classification} for three examples of instances from the Government Actions Tracker database. The \textit{Context} feature was omitted here because it is simply the concatenation of the second and third columns.}
    \label{tab:examples}
    \begin{tabularx}{\textwidth}{ | m{1.82cm} | m{3cm} | m{4.7cm} | m{2.2cm} | }
    \hline
    \textbf{Theme} & \textbf{Action} & \textbf{Circumstance} & \textbf{Classification} \\
    \hline
    Environment & \textit{Estabelece os proce-dimentos administrativos no âmbito do Ibama para a de-legação de licenciamento ambiental de competência (...) ou Municipal.} & \textit{Autoriza o IBAMA a repassar para Estados e municípios qualquer processo de licenciamento ambiental de sua responsabilidade, incluindo empreendimentos em terras indígenas, áreas protegidas e na costa brasileira. Entende-se como desregulação porque (...) Lei Complementar nº 140/2011 (Art. 2º § 2º).} & Deregulation \\
    \hline
    Institutional & \textit{Revoga atos normativos (...) Fundação Nacional do Índio - Funai, conforme Decreto (...).} & \textit{Declara revogados os atos normativos da Procuradoria Federal Especializada junto à Fundação Nacional do Índio: Ordem de Serviço (...) setembro de 2008, p. 5-7.} & Revocation \\
    \hline
    Energy & \textit{Recomenda (...) de energia no âmbito do Programa de Parcerias de Investimentos.} & \textit{Opina favoravelmente e submeter (...) no âmbito do Programa de Parcerias de Investimentos - PPI (...) no ano de 2021.} & Privatization \\
    \hline
    \end{tabularx}
\end{table}

\begin{figure*}[ht]
\vskip 0.2in
\begin{center}
\centerline{\includegraphics[width=1.0\textwidth]{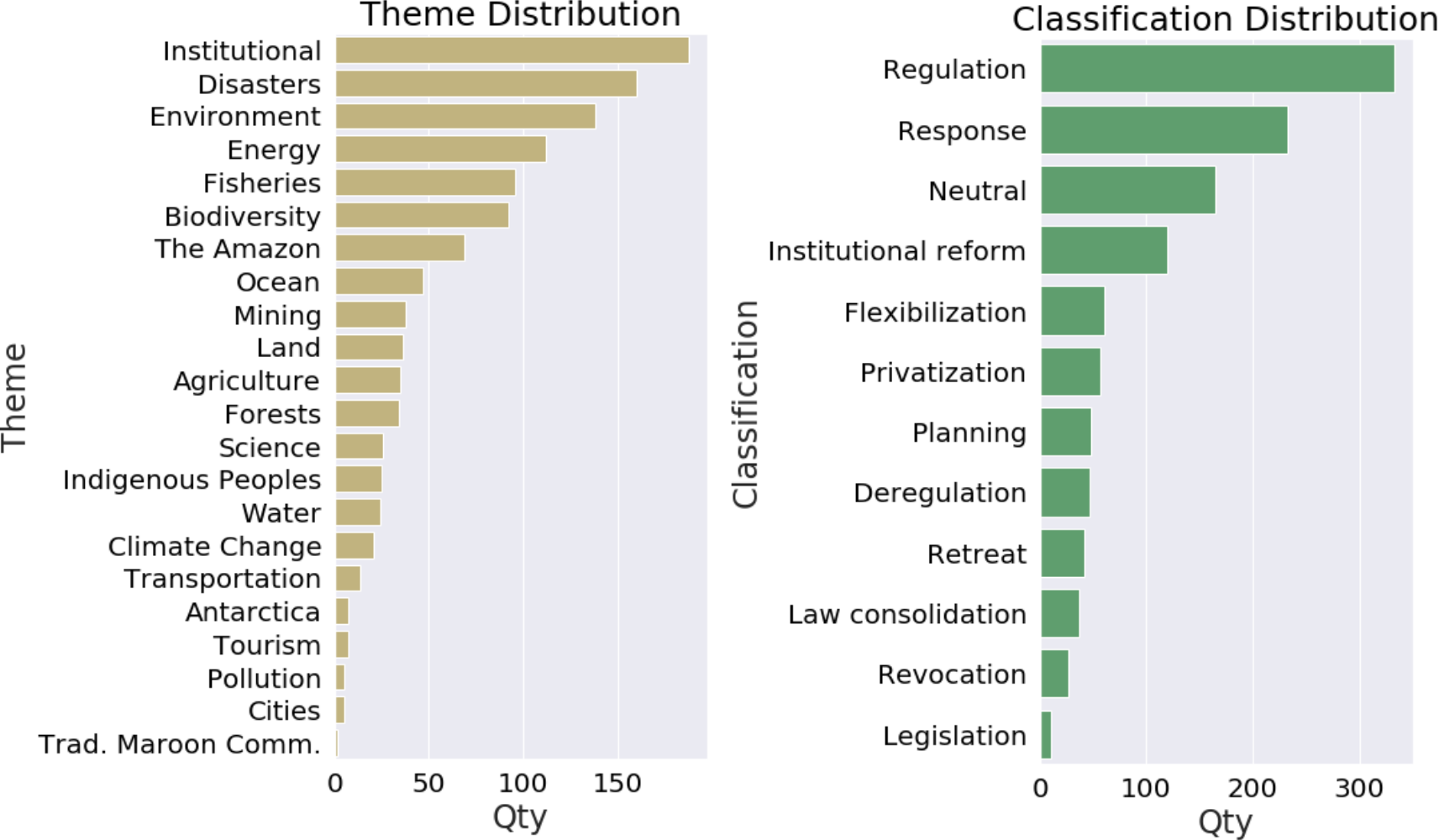}}
\caption{Distribution of Theme and Classification variables in the GAT dataset we used.}
\label{gat_distr}
\end{center}
\vskip -0.2in
\end{figure*}

After the human supervision stage, the verified and enriched data are sent to a separate database, the \textit{Government Actions Tracker} (GAT) database. Table~\ref{tab:examples} shows the \textbf{Theme}, \textbf{Action}, \textbf{Circumstance} columns and their respective \textbf{Classification} column for three examples of instances, obtained from the GAT database (held in the original language). The version of GAT dataset we used in this work has 1,181 instances and no missing data in the Theme or Classification variables. Figure \ref{gat_distr} shows their distributions in the dataset used. The \textit{Action} feature has $29.1 \pm 19.6$ words on average; the \textit{Circumstance} one, $70.0 \pm 54.0$ words. The first record dates from January 1, 2019; the last, July 12, 2021. 

Due to the small size of the GAT database, the training with the original 12 classes proved to be very unstable, since the objective was to predict the \textbf{Classification} of a document given only its \textbf{Context}. Thus, we regrouped the previous classes of the \textbf{Classification} variable into three major classes, as follows:

\begin{itemize}
    \item \textbf{Regulation:} Regulation, Planning and Response;
    \item \textbf{Neutral:} Neutral, Retreat and Legislation;
    \item \textbf{Deregulation:} Privatization, Deregulation, Flexibilization, Institutional reform, Law consolidation and Revocation.
\end{itemize}

Therefore, the dataset ended up with the following proportions: Regulation (52.0\%), Neutral (18.5\%) and Deregulation (29.5\%). 

\subsection{Experiment Description}
In order to discover the most suitable model for classifying the Acts in the BFOG, we built and compared four models. We start with two simpler models, based on Naive Bayes \cite{Zhang2005} and Bidirectional LSTMs (BiLSTMs) \cite{schuster1997bidirectional}. We also fine-tuned a BERTimbau Base model \cite{Souza2020} (we call ``BERT1") directly on the classification task of GAT. Finally, our fourth model (``BERT2") was also based on the BERTimbau Base model, but in this case we performed two subsequential fine-tunings: first on 500k documents from the entire (and unprocessed) BFOG database since 2002\footnote{For the Executive branch, it is available at: \url{https://www.politicaporinteiro.org/base-de-atos-do-executivo/}.} (the blue database in Figure \ref{data_pipeline}), and then on the GAT database to classify the government acts.

All models were trained under a stratified cross-validation with \textit{k-folds = 10}. We also performed different strategies for data augmentation with sequence-to-sequence models, such as T5 \cite{JMLR:v21:20-074} and Pegasus \cite{zhang2020pegasus}. However, as none of them outperformed the models without data augmentation, we left them out of the scope of this work. Further, we describe our four models in more details below:

\subsubsection{Simpler models: Naive Bayes and BiLSTM}
As our first models, we trained a Multinomial Naive Bayes (NB) model and a bidirectional LSTM (BiLSTM) network with 32 LSTM units, with batch size of 32, maximum sequence length of 50 for 4 epochs - a Gridsearch was performed to select these hyperparameters. The net weights were initialized from a previously trained matrix (CBOW) of 300 dimensions \cite{hartmann2017}.

\subsubsection{BERT1}
We also fine-tuned a BERTimbau Base model, a pretrained BERT model in Portuguese with 12 layers and 110 million parameters, on the GAT classification task. After multiple inspections, we trained the model for 15 epochs, with 512 as the maximum sequence length, a learning rate of 5e-5, a batch size of 128 and an Adam's optimizer algorithm with weight decay.

\subsubsection{BERT2}
As Table~\ref{tab:examples} illustrates, the sentences in the BFOG have a very characteristic writing style, with many technical terms and legal jargon. Hence, instead of fine-tuning BERTimbau Base directly on the GAT classification task, we first trained it on a Masked Language Model (MLM) task on a dataset of 500k documents, since 2002, from the BFOG, scraped from the official websites. We only use the body of the documents (the ``\textit{ementa detalhada}"), with minimal text processing -- basically the removal of special characters like ``$\backslash$n" or ``$\backslash$t". The final file was 1.5GB of plain text and we trained the MLM model for 10 epochs. Finally, the model was then fine-tuned in the classification problem of the GAT dataset, as in the previous case and with the same hyperparameters as BERT1, as both responded in a similar way to the tests.

\section{Results}

Table~\ref{tab:final_results} summarizes the results obtained for each model. Pre-trained models were superior to baselines in all metrics, and BERT2 performed slightly above BERT1, although the difference is not statistically relevant. Among the baseline models, even with the LSTM network having been initialized with an already trained weight matrix and after a careful choice of parameters, its results in general were still inferior to the NB, which reinforces the idea that, without pre-trained models based on the Transformer architecture, recurrent networks can be quite limited in offering a substantial improvement over simpler models.

Considering the proportions of each class in each experiment, one can see that specially the first two models, BERT1 and BERT2, are promising, despite the challenges of dealing with a small database for the current standards of the state-of-the-art pretrained models and also despite of the imbalanced classes. 

\begin{table}
\centering
\caption{Summary of the Matthews Correlation Coefficient (MCC), Accuracy (Acc) and weighted F1-score for each model in the task of classifying the \textit{Context} information into one of the three classes: \textit{Regulation}, \textit{Neutral} or \textit{Deregulation}. Results were cross-validated with \textit{k-fold = 10}. In bold, the best ones.}
\label{tab:final_results}
\begin{tabular}{| c | c | c | c | }
\hline
\textbf{Model} & \textbf{MCC} & \textbf{Acc} & \textbf{F1-score} \\
\hline
\textbf{BERT2} & $\mathbf{0.538 \pm 0.046}$ & $\mathbf{0.725 \pm 0.027}$ & $\mathbf{0.714 \pm 0.031}$  \\
\hline
\textbf{BERT1} & $0.530 \pm 0.050$ & $0.716 \pm 0.031$ & $0.710 \pm 0.032$ \\
\hline
\textbf{BiLSTM (CBOW 300)} & $0.401 \pm 0.090$ & $0.644 \pm 0.055$ & $0.618 \pm 0.059$ \\
\hline
\textbf{Naive Bayes} & $0.429 \pm 0.032$ & $0.658 \pm 0.019$ & $0.601 \pm 0.030$ \\
\hline
\end{tabular}
\end{table}

\section{Conclusion and Future Work}
In this work, we present strategy to leverage automated techniques and expert knowledge to track and classify potentially harmful changes in environmental policies directly from texts in Brazilian Federal Official Gazette. In addition to this strategy, we contribute with the Government Actions Tracker, a new challenging curated dataset, in Portuguese, of federal government acts related to Brazilian environmental policies. Also, we designed and compared four different NLP models on the classification task posed by this dataset, from the simpler models to the state-of-the-art ones. Monitoring each act published by the government in order to inform civil society is an extremely challenging task and there is still no single practical solution to solve it. While a rule-based system working jointly with domain experts already deliver immeasurable value when it comes to policy monitoring, it is also paramount that the latest NLP technologies be considered to increase the scalability and performance of these systems. Hence, among the future work are the expansion of the annotated datasets, as well as the improvement of the best models presented here so that they can keep the quality and stability of the classification for a greater number of classes.

\section{Acknowledgements}
This work is the result of an academic partnership between the Escola Politécnica of Universidade de São Paulo (EP-USP) and \textit{Política Por Inteiro}. Without the data curation efforts of Política Por Inteiro experts, building these NLP models would not have been possible. Also, this work was financed in part by the \textit{Coordenação de Aperfeiçoamento de Pessoal de Nível Superior} (CAPES, Finance Code 001), the \textit{Itaú Unibanco S.A.}, through the \textit{Programa de Bolsas Itaú} (PBI) of the Centro de Ciência de Dados (C2D) of EP-USP, and by the \textit{Conselho Nacional de Desenvolvimento Científico e Tecnológico} (CNPq) (grant 310085/2020-9). We also thank the \textit{Center for Artificial Intelligence} (C4AI-USP), with support by the \textit{Fundação de Amparo à Pesquisa do Estado de São Paulo} (FAPESP, grant 2019/07665-4) and by the \textit{IBM Corporation}. The data, views and opinions expressed in this article are those of the authors and do not necessarily reflect the official policy or position of the financiers.

%
%
%
\bibliographystyle{splncs04}
\bibliography{references}

\begin{thebibliography}{10}
\providecommand{\url}[1]{\texttt{#1}}
\providecommand{\urlprefix}{URL }
\providecommand{\doi}[1]{https://doi.org/#1}

\bibitem{Paschoal2021}
{Andr{\'{e}} F. A. Paschoal, Paulo Pirozelli, Valdinei Freire, Karina V.
  Delgado, Sarajane M. Peres, Marcos M. Jos{\'{e}}, Fl{\'{a}}vio N.
  Ca{\c{c}}{\~{a}}o, Andr{\'{e}} S. Oliveira, Anarosa A. F. Brand{\~{a}}o, and
  Anna H. R. Costa}, F.G.C.: {Pir{\'{a}}: A Bilingual Portuguese-English
  Dataset for Question-Answering about the Ocean}. In: 30th ACM International
  Conference on Information and Knowledge Management (CIKM'21) (2021).
  \doi{https://doi.org/10.1145/3459637.3482012}

\bibitem{caccao2021deepage}
Ca{\c{c}}{\~a}o, F.N., Jos{\'e}, M.M., Oliveira, A.S., Spindola, S., Costa,
  A.H.R., Cozman, F.G.: Deepag{\'e}: Answering questions in portuguese about
  the brazilian environment. In: Brazilian Conference on Intelligent Systems.
  pp. 419--433. Springer (2021)

\bibitem{Esalq2020}
CEPEA-ESALQ: {PIB do Agroneg{\'{o}}cio Brasileiro - Centro de Estudos
  Avan{\c{c}}ados em Economia Aplicada - CEPEA-Esalq/USP}  (2021)

\bibitem{Devlin2018}
Devlin, J., Chang, M.W., Lee, K., Toutanova, K.: {BERT: Pre-training of deep
  bidirectional transformers for language understanding}. In: Proceedings of
  the 2019 Conference of the North. vol.~1, pp. 4171--4186. Association for
  Computational Linguistics, Stroudsburg, PA, USA (2019).
  \doi{10.18653/v1/N19-1423}, \url{http://aclweb.org/anthology/N19-1423}

\bibitem{ferraz2021debacer}
Ferraz, T., Alcoforado, A., Bustos, E., Oliveira, A., Gerber, R., Müller, N.,
  d’Almeida, A., Veloso, B., Costa, A.: Debacer: a method for slicing
  moderated debates. In: Anais do XVIII Encontro Nacional de Inteligência
  Artificial e Computacional. pp. 667--678. SBC, Porto Alegre, RS, Brasil
  (2021). \doi{10.5753/eniac.2021.18293},
  \url{https://sol.sbc.org.br/index.php/eniac/article/view/18293}

\bibitem{Grimmer2013}
Grimmer, J., Stewart, B.M.: {Text as data: The promise and pitfalls of
  automatic content analysis methods for political texts}. Political Analysis
  \textbf{21}(3),  267--297 (2013). \doi{10.1093/pan/mps028}

\bibitem{hartmann2017}
Hartmann, N.S., Fonseca, E.R., Shulby, C.D., Treviso, M.V., Rodrigues, J.S.,
  Aluísio, S.M.: Portuguese word embeddings: Evaluating on word analogies and
  natural language tasks. In: Anais do XI Simpósio Brasileiro de Tecnologia da
  Informação e da Linguagem Humana. pp. 122--131. SBC, Porto Alegre, RS,
  Brasil (2017), \url{https://sol.sbc.org.br/index.php/stil/article/view/4008}

\bibitem{hatonen2021from}
Hätönen, V., Melzer, F.: From talk to action with accountability: Monitoring
  the public discussion of policy makers with deep neural networks and topic
  modelling (2021), \url{https://www.climatechange.ai/papers/icml2021/75}

\bibitem{liu1907roberta}
Liu, Y., Ott, M., Goyal, N., Du, J., Joshi, M., Chen, D., Levy, O., Lewis, M.,
  Zettlemoyer, L., Stoyanov, V.: Roberta: A robustly optimized bert pretraining
  approach. corr abs/1907.11692 (2019). URL: http://arxiv. org/abs/1907.11692
  (1907)

\bibitem{Luccioni2020}
Luccioni, A., Baylor, E., Duchene, N.: {Analyzing Sustainability Reports Using
  Natural Language Processing}  (2020), \url{http://arxiv.org/abs/2011.08073}

\bibitem{JMLR:v21:20-074}
Raffel, C., Shazeer, N., Roberts, A., Lee, K., Narang, S., Matena, M., Zhou,
  Y., Li, W., Liu, P.J.: {Exploring the Limits of Transfer Learning with a
  Unified Text-to-Text Transformer}. Journal of Machine Learning Research
  \textbf{21}(140),  1--67 (2020), \url{http://jmlr.org/papers/v21/20-074.html}

\bibitem{Rochedo2018}
Rochedo, P.R.R., Soares-Filho, B., Schaeffer, R., Viola, E., Szklo, A., Lucena,
  A.F.P., Koberle, A., Davis, J.L., Raj{\~{a}}o, R., Rathmann, R.: {The threat
  of political bargaining to climate mitigation in Brazil}. Nature Climate
  Change  \textbf{8}(8),  695--698 (aug 2018). \doi{10.1038/s41558-018-0213-y},
  \url{http://www.nature.com/articles/s41558-018-0213-y}

\bibitem{Rolnick2019}
Rolnick, D., Donti, P.L., Kaack, L.H., Kochanski, K., Lacoste, A., Sankaran,
  K., Ross, A.S., Milojevic-Dupont, N., Jaques, N., Waldman-Brown, A.,
  Luccioni, A., Maharaj, T., Sherwin, E.D., {Karthik Mukkavilli}, S., Kording,
  K.P., Gomes, C., Ng, A.Y., Hassabis, D., Platt, J.C., Creutzig, F., Chayes,
  J., Bengio, Y.: {Tackling climate change with machine learning}. arXiv
  (2019)

\bibitem{schuster1997bidirectional}
Schuster, M., Paliwal, K.K.: Bidirectional recurrent neural networks. IEEE
  transactions on Signal Processing  \textbf{45}(11),  2673--2681 (1997)

\bibitem{SilvaJunior2021}
{Silva Junior}, C.H., Pess{\^{o}}a, A.C., Carvalho, N.S., Reis, J.B., Anderson,
  L.O., Arag{\~{a}}o, L.E.: {The Brazilian Amazon deforestation rate in 2020 is
  the greatest of the decade}. Nature Ecology and Evolution  \textbf{5}(2),
  144--145 (2021). \doi{10.1038/s41559-020-01368-x}

\bibitem{Souza2020}
Souza, F., Nogueira, R., Lotufo, R.: {BERTimbau: Pretrained BERT Models for
  Brazilian Portuguese}. In: Lecture Notes in Computer Science (including
  subseries Lecture Notes in Artificial Intelligence and Lecture Notes in
  Bioinformatics), vol. 12319 LNAI, pp. 403--417 (2020).
  \doi{10.1007/978-3-030-61377-8\_28}

\bibitem{Reuter2020}
Spring, J.: {Deforestation boosts Brazil greenhouse gas emissions as global
  emissions fall | Reuters} (2020),
  \url{https://www.reuters.com/article/us-brazil-environment-emissions-idUSKBN22X2AA}

\bibitem{stede2021climate}
Stede, M., Patz, R.: The climate change debate and natural language processing.
  In: Proceedings of the 1st Workshop on NLP for Positive Impact. pp. 8--18
  (2021)

\bibitem{Vaswani2017}
Vaswani, A., Shazeer, N., Parmar, N., Uszkoreit, J., Jones, L., Gomez, A.N.,
  Kaiser, {\L}., Polosukhin, I.: {Attention is all you need}. Advances in
  Neural Information Processing Systems  \textbf{2017-Decem}(Nips),  5999--6009
  (2017)

\bibitem{West2019}
West, T.A.P., B{\"{o}}rner, J., Fearnside, P.M.: {Climatic Benefits From the
  2006–2017 Avoided Deforestation in Amazonian Brazil}. Frontiers in Forests
  and Global Change  \textbf{2}(September),  1--11 (2019).
  \doi{10.3389/ffgc.2019.00052}

\bibitem{Zhang2005}
Zhang, H.: {Exploring conditions for the optimality of na{\"{i}}ve bayes}.
  International Journal of Pattern Recognition and Artificial Intelligence
  \textbf{19}(2),  183--198 (2005). \doi{10.1142/S0218001405003983}

\bibitem{zhang2020pegasus}
Zhang, J., Zhao, Y., Saleh, M., Liu, P.: Pegasus: Pre-training with extracted
  gap-sentences for abstractive summarization. In: International Conference on
  Machine Learning. pp. 11328--11339. PMLR (2020)

\end{thebibliography}

\end{document}